\documentclass[aps,showpacs,amsmath,amssymb,floatfix,
twocolumn]{revtex4-1}
\usepackage{graphicx}
\usepackage{dcolumn}
\usepackage{bm}
\usepackage{hyperref}
\hypersetup{linkcolor=red,citecolor=blue,urlcolor=cyan}
\begin{document}
\title{Matching slowly-rotating spacetimes split by dynamic thin shells
}

\author{Jonas P. Pereira$^{1,2}$ and Jorge A. Rueda$^{3,4,5,6,7}$}
\email{e-mails to: jpereira@camk.edu.pl, jorge.rueda@icra.it}
\affiliation{$^{1}$ Núcleo de Astrofísica e Cosmologia (Cosmo-Ufes) \& Departamento de Física, Universidade Federal do Espírito Santo, Vitória, 29075-910, ES, Brazil}
\affiliation{$^{2}$Nicolaus Copernicus Astronomical Center, Polish Academy of Sciences, Warsaw, 00-716, Poland}
\affiliation{$^{3}$ICRA, Dipartimento di Fisica, Sapienza Universit\`a  di Roma, Piazzale Aldo Moro 5, I-00185 Roma, Italy}
\affiliation{$^{4}$ICRANet, Piazza della Repubblica 10, I-65122 Pescara, Italy}
\affiliation{$^{5}$ICRANet-Ferrara, Dipartimento di Fisica e Scienze della Terra, Universit\`a degli Studi di Ferrara, Via Saragat 1, I--44122 Ferrara, Italy}
\affiliation{$^{6}$Dipartimento di Fisica e Scienze della Terra, Universit\`a degli Studi di Ferrara, Via Saragat 1, I--44122 Ferrara, Italy}
\affiliation{$^{7}$INAF, Istituto di Astrofisica e Planetologia Spaziali, Via Fosso del Cavaliere 100, 00133 Rome, Italy}

\date{\today}

\begin{abstract}
We investigate within the Darmois-Israel thin shell formalism the match of neutral and asymptotically flat, slowly rotating spacetimes (up to the second order in the rotation parameter) when their boundaries are dynamic. It has several important applications in general relativistic systems, such as black holes and neutron stars, which we exemplify. We mostly focus on stability aspects of slowly rotating thin shells in equilibrium and surface degrees of freedom on the hypersurfaces splitting the matched slowly rotating spacetimes, e.g., surface energy density and surface tension. We show that the stability upon perturbations in the spherically symmetric case automatically implies stability in the slow rotation case. In addition, we show that when matching slowly rotating Kerr spacetimes through thin shells in equilibrium, surface degrees of freedom can decrease compared to their Schwarzschild counterparts, meaning that energy conditions could be weakened. Frame-dragging aspects of the match of slowly rotating spacetimes are also briefly discussed.

\end{abstract}

\maketitle

\section{Introduction}

Although matching two spherically symmetric spacetimes is a simple task (see \cite{2005CQGra..22.4869L}, for instance), this is not the case for axially symmetric ones \cite{1996CQGra..13.1885M}. Nevertheless, particular glues have already been applied to several scenarios and contexts in the slow rotation case. We mention, for instance, the collapse of slowly rotating thin shells (onto already formed black holes \cite{1968PhRv..170.1187D}, generating slowly rotating black holes \cite{1978PhRvD..18.1020K} or associated with regular interior spacetimes \cite{PhysRevD.90.064042}), the analysis of kinematical effects slowly rotating spacetime glues should exhibit \cite{1986CQGra...3..335P, 1978PhRvD..18.1757O},
and scalar fields non-minimally coupled to glued spacetimes \cite{2014PhRvD..90d4053M}. Concerning the match of slowly rotating spacetimes up to the second order in the rotation parameter, the work of \citet{1968PhRv..170.1187D} is noteworthy. They did it for the first time by joining the Minkowski and the Kerr spacetimes using prescribed axially symmetric hypersurfaces in equilibrium. It was shown that there are infinite ways of joining these spacetimes, each characterized by a spinning shell with a different degree of ellipticity. In addition, rigid rotation of observers inside the shell with the shell itself does not occur generally. It was not the interest of that work to assess the stability of the equilibrium radii.

Another more recent example of interesting glue is the work of \citet{PhysRevD.90.064042}, who investigated the matching of a slowly rotating Kerr-Newman spacetime with the de Sitter one using a thin-shell in equilibrium (its stability was not investigated either). As expected, they also found that the surface degrees of freedom are polar angle-dependent on the thin shell. Cases were found where the surface energy density could be negative, but the authors deemed them unphysical. For further interesting slowly rotating thin-shell applications, see \citep{2015PhRvD..92l4030P,2016CQGra..33b5005U,2016PhRvD..94f4015U,2022PhRvD.105b4002B,2022ForPh..70f0177A,2023ForPh..71b0108T} and references therein.

Clear assets of matching axially symmetric spacetimes are the possibility of having manifolds without singularities and the possibility of more realistic models for astrophysical systems. In addition, the dragging of inertial frames is inherent in these spacetimes. We show here how this kinematic effect would allow us to scrutinize layered systems in astrophysics, shedding some light on their constitution.

We address the problem of matching neutral, arbitrary, slowly rotating Hartle's spacetimes split by dynamic hypersurfaces, examining some of their subtleties. To the best of our knowledge, this has not been done before.  We limit our analysis to the second order in the rotation parameter since it is not known how to generically glue two axially symmetric spacetimes with arbitrary rotation parameters. As a byproduct of our investigation, we show, for instance, that the stability of thin shells in the spherically symmetric case automatically implies the stability of thin shells when slow rotation (up to the second order in the rotation parameter) is present. In addition, we show that there are (realistic) situations where the corrections to the surface energy density and surface tension in the spherically symmetric case are negative. This seems to be of interest, for it would point to the likelihood of violating some energy conditions in some regions of surfaces of discontinuity in the non-perturbative context.

This article is organized as follows. In the next section, we deal with the problem of matching two slowly rotating Hartle's spacetimes split by a dynamic hypersurface. Section \ref{Rconst} is devoted to the solution of the established system of equations for static and away-from-equilibrium hypersurfaces. The generalizations of the surface energy density and surface tension in the case of slow rotation are presented in Section \ref{sigmarot}. Kinematic effects associated with our generic glues are addressed in Section \ref{drag}. In Section \ref{KN}, we apply our formalism to the important case of matching two Kerr spacetimes and investigate the behavior of surface quantities and some aspects of the energy conditions. A summary and many astrophysical applications of our work can be found in section \ref{conc}.

We use geometric units throughout the article and the metric signature $-2$.

\section{Matching slowly rotating spacetimes}
\label{Hartle}

In the slowly rotating case, following \citet{1967ApJ...150.1005H}, we assume that each spacetime that we glue can be written as
\begin{eqnarray}
&&ds^2=e^{\nu(r)}[1+2\epsilon^2h(r,\theta)]dt^2-e^{\lambda(r)}[1+2\epsilon^2j(r,\theta)]dr^2\nonumber \\ &-& r^2[1+2\epsilon^2k(r,\theta)](d\theta^2+\sin^2\theta\{d\varphi-\epsilon\omega(r) dt \}^2)\label{slowrotmetric},
\end{eqnarray}
where
\begin{eqnarray}
h(r, {\theta})&=&h_0(r)+ h_2(r)P_2(\cos{\theta})\label{hdecomp},\\
j(r, {\theta})&=&j_0(r)+ j_2(r)P_2(\cos{\theta})\label{jdecomp},\\
k(r, {\theta})&=&k_2(r)P_2(\cos{\theta})\label{kdecomp}.
\end{eqnarray}
As usual, we are defining $P_2(\cos{\theta})$ as the second order Legendre polynomial [$P_2(\cos{\theta})=(3\cos^2{\theta}-1)/2$].

In these approximations, we go up to the \emph{second order in the rotation parameter}, fingerprinted by the dummy (dimensionless) quantity ``$\epsilon$''. It was put in Eq. (\ref{slowrotmetric}) just as a mere indicator of the order of the rotational expansion
taken into account. The functions $h,j,k,\omega$ as given by Eqs. (\ref{hdecomp}--\ref{kdecomp}) are to be found by Einstein's equations in the scope of axially symmetric solutions (see Ref.~\cite{1967ApJ...150.1005H} for the associated equations, some solutions, and properties). The background spacetime that is the seed for the Hartle metric is a spherically symmetric one (for the case of a black hole spacetime, it would be the Schwarzschild metric), as can be seen by putting $\epsilon=0$ in Eq. \eqref{slowrotmetric}.

To match two spacetimes (that can then be used to obtain spacetimes made up of an arbitrary number of glues), one must also know their common hypersurface (that is to say, it must be well-defined), as well as how the coordinates of each manifold are related to the ones defined on such a hypersurface. Once this is obtained, finding the surface energy-momentum tensor that would allow such a glue is more of an algebraic exercise, made possible mainly by Lanczos' and Israel's seminal works \cite{1924AnP...379..518L, 1966NCimB..44....1I,  2005CQGra..22.4869L, 2004reto.book.....P}.

To begin with, let us choose the intrinsic coordinates of the splitting hypersurface, defined as $\Sigma$, as
$y^a=(\tilde\tau,\tilde{\theta},\tilde{\varphi})$. For now, $\tilde\tau$ is just a label for a time-like coordinate on $\Sigma$.
Further ahead, such a choice will be justified when we try to relate it with the proper time measured by observers on $\Sigma$,
at least to some order in ``$\epsilon$''. We shall assume that the equation for $\Sigma$ is given by
\begin{equation}
\Psi(r,\tilde\tau,\tilde{\theta})=0,\quad \Psi(r,\tilde\tau,\tilde{\theta})=r- R(\tilde\tau) - \epsilon^2B(\tilde\tau, \tilde{\theta})\label{hypersurfeq}.
\end{equation}
Up to this point, $R$ and $B$ are unknown functions, and ``$\epsilon$'' indicates the rotational order of the factors involved.
In addition, let us assume that the spacetime coordinates relate to the ones of the hypersurface as
\begin{equation}
t=T(\tilde\tau)+\epsilon^2A(\tilde\tau,\tilde{\theta}),\;\; \theta=\tilde{\theta},\;\; d\varphi=d\tilde{\varphi}+\epsilon C(\tilde\tau)d\tilde\tau \label{coordtrans}.
\end{equation}
Just not to overload the notation, we dropped the ``$\pm$'' indexes that should be present in the coordinates of the spacetimes to be matched. Such labels would be related to a region above (``$+$'') and below (``$-$'') $\Sigma$, naturally defined by its normal vector.

With the Ansatz given by Eqs.~(\ref{hypersurfeq}) and (\ref{coordtrans}), our task is to find the functions $A$, $B$, $C$, $R$, $T$ that lead the metrics given
by Eq.~(\ref{slowrotmetric}) to be continuous when projected onto $\Sigma$  (continuity of the first fundamental form). This is the first junction condition to be imposed when matching
spacetimes whose resultant one leads to a distributional solution to Einstein's equations \cite{2004reto.book.....P}. From Eqs.~(\ref{slowrotmetric}) and (\ref{coordtrans}), we have to impose that (for any $F$, $\dot{F}\equiv \partial F/ \partial \tilde{\tau}$)
\begin{equation}
C=\omega(R)\dot{T}\label{cvalue}
\end{equation}
and
\begin{equation}
e^{\nu(R)}\dot{T}^2-e^{\lambda(R)}\dot{R}^2=1\label{sphersymcond},
\end{equation}
to eliminate the first-order terms in the rotational parameter of the induced metric and retrieve our spherically symmetric solution with $\tilde\tau$ as the proper time on $\Sigma$.
From Eq.~(\ref{sphersymcond}), we see that the well-definiteness of the induced metric does not constrain $R(\tilde\tau)$; it is a free function whose dynamics will be related to the spherically symmetric configuration. We know from such a case that its determination is just possible when an equation of state for the {(perfect-like)} fluid on $\Sigma$ is given \cite{2005CQGra..22.4869L,2011PhRvD..83j4009E,PhysRevD.90.123011}.

Since we are working perturbatively in ``$\epsilon$'', we can expand the functions $A$ and $B$ up to the second-order Legendre polynomials
in the same fashion as it has been done previously for $h,j,k$. Therefore, let us assume
\begin{equation}
B(\tilde\tau, \tilde{\theta})= B_0(\tilde{\tau})+ B_2(\tilde\tau)P_2(\cos\tilde{\theta})\label{Bdecomp},
\end{equation}
and
\begin{equation}
A(\tilde\tau, \tilde{\theta})=A_0(\tilde\tau) + A_2(\tilde\tau)P_2(\cos\tilde{\theta})\label{Adecomp}.
\end{equation}
On the hypersurface, $\Sigma$, the functions $h_0$, $j_0$, $h_2$, $j_2$ and $k_2$ [see Eqs.~(\ref{hdecomp}--\ref{kdecomp})] have their radial dependence just replaced by $R$, given that they are already second order functions in the rotational parameter. Notwithstanding, the spherically symmetric function $e^{\nu}$ on the hypersurface $\Sigma$, on account of Eqs.~(\ref{hypersurfeq})
and (\ref{Bdecomp}), changes to
\begin{equation}
e^{\nu(r)}\approx e^{\nu(R)}\left(1+\frac{\partial \nu}{\partial R}\epsilon^2B \right)\label{g00decomp},
\end{equation}
similarly to $e^{\lambda}$. Therefore, further corrections appear when one works up to $\epsilon^2$. The one given by Eq.~(\ref{g00decomp}) is
due to the shape change of $\Sigma$ due to rotation.

From Eqs.~(\ref{slowrotmetric}), (\ref{coordtrans}), (\ref{cvalue}), (\ref{sphersymcond}) and (\ref{g00decomp}), the induced metric on $\Sigma$ can be cast as
\begin{widetext}
\begin{eqnarray}
ds^2_{\Sigma}&=&d\tilde{\tau}^2 \left\{1 + \epsilon	^2\left[ e^{\nu(R)}\left(2\dot{T}\dot{A}+ 2h\dot{T}^2+\frac{\partial \nu}{\partial R}B\dot{T}^2 \right) - e^{\lambda(R)} \left( 2\dot{R}\dot{B}+ 2j\dot{R}^2+\frac{\partial \lambda}{\partial R}B\dot{R}^2 \right)\right] \right\} \nonumber
\\
&+& 2\epsilon^2\left( e^{\nu(R)}\dot{T}\frac{\partial A}{\partial \tilde{\theta}}- e^{\lambda(R)}\dot{R}\frac{\partial B}{\partial \tilde{\theta}}\right)d\tilde\tau d\tilde{\theta} - (R^2+2RB+2kR^2)(d\tilde{\theta}^2+\sin\tilde{\theta}^2d\tilde{\varphi}^2)\label{inducmetricrude}.
\end{eqnarray}
\end{widetext}
If we now substitute Eqs.~(\ref{hdecomp}), (\ref{jdecomp}), (\ref{kdecomp}), (\ref{Bdecomp}) and (\ref{Adecomp}) into Eq.~(\ref{inducmetricrude}), the well-definiteness of the induced metric on $\Sigma$ is guaranteed by means of the following conditions:
\begin{eqnarray}
&&e^{\nu(R)}\left(\dot{T}\dot{A}_0 + h_0\dot{T}^2 \right)-e^{\lambda(R)}\left(\dot{R}\dot{B}_0 + j_0\dot{R}^2 \right) +\nonumber\\ &&\frac{B_0}{2}\left(e^{\nu(R)}\dot{T}^2\frac{\partial \nu}{\partial R} - e^{\lambda(R)}\dot{R}^2\frac{\partial \lambda}{\partial R} \right)=0\label{A0cond},\\
&&e^{\nu(R)} \dot{T}A_2 = e^{\lambda(R)}\dot{R}B_2\label{A2cond},\\
&&\left[ B_2 +k_2R\right]^+_-=0 \label{B2cond},\\
&&[B_0]^+_-=0\label{B0cond},\\
&&[\alpha_2]^+_-=0\label{alpha2cond},
\end{eqnarray}
with
\begin{eqnarray}
\alpha_2 &\doteq& e^{\nu(R)}\left(\dot{T} \dot{A_2} + h_2 \dot{T}^2 + \frac{1}{2}\frac{\partial \nu}{\partial R} B_2 \dot{T}^2 \right)\nonumber\\ &-& e^{\lambda(R)}\left(\dot{R} \dot{B_2} + j_2 \dot{R}^2 + \frac{1}{2}\frac{\partial \lambda}{\partial R} B_2 \dot{R}^2 \right) \label{alpha2}.
\end{eqnarray}
In the above equations, we have defined $[A]^+_- \doteq A^{+}|_{\Sigma}- A^{-}|_{\Sigma}$ as the \textit{jump of $A$ across $\Sigma$}. We stress that each of Eqs.~(\ref{A0cond}) and (\ref{A2cond}) are two equations, related to each region ($\pm$) defined by $\Sigma$. We then have
eight unknown functions, $(A_0^{\pm}, A_2^{\pm},B_2^{\pm}, B_0^{\pm})$, to seven equations, Eqs.~(\ref{A0cond}--\ref{alpha2cond}). The
missing equation is related to the freedom in deforming the shape of the surface, splitting the two glued regions of space for a fixed eccentricity. 
Note that $B_0^{\pm}=0$ (or any given constant) renders Eq.~(\ref{B0cond}) an identity, and in this case, we are left with a system of six equations to six variables. Equations~(\ref{A0cond}--\ref{alpha2}) allow us to express the (continuous) induced metric on $\Sigma$ in a simplified way, namely,
\begin{eqnarray}
ds^2_{\Sigma} &&= d\tilde{\tau}^2 \big[1+ 2\epsilon^2 \{ \alpha_2 \}^+_-  P_2(\cos\tilde\theta) \big] - \big[R^2 + 2\epsilon^2 R\{ B_0\}^+_-\nonumber \\ &&  + 2\epsilon^2 R\{ B_2+k_2R \}^+_- P_2(\cos\tilde\theta)\big] (d\tilde\theta^2+\sin\tilde\theta^2 d \tilde\varphi^2)\label{inducmetricfinal},
\end{eqnarray}
where we are defining for a given quantity $F$ {on $\Sigma$}, $\{ F \}^+_-\equiv (F|_{\Sigma^+} + F|_{\Sigma^-})/2$. Of course, that simplified way of expressing the induced metric is due to our coordinate freedom on $\Sigma$. 

We stress that all of our previous reasoning remains the same if we perform the change $P_2(\cos\theta)\rightarrow {\cal C}P_2(\cos\theta)$, with ${\cal C}$ an arbitrary constant. This is related to the freedom we always have in choosing the ellipticity of $\Sigma$ when matching two slowly rotating spacetimes.
In addition, $\tilde\tau$ is not the proper-time on $\Sigma$ if one takes into account Eqs.~(\ref{A0cond}--\ref{alpha2}). This is not a problem since the coordinates on $\Sigma$ can be chosen freely. Finally, we recall that the thin-shell formalism leads to surface quantities that are independent of the coordinate systems of the glued spacetimes and the hypersurface. We chose Hartle's spacetime and $y^a$ because, respectively, it is physically appealing and convenient for slowly rotating extended systems such as neutron stars, and it leads to a simple induced metric on $\Sigma$.

\section{Static and away-from-equilibrium thin shells}
\label{Rconst}

We solve Eqs.~(\ref{A0cond}--\ref{alpha2}) for the case $\Sigma$ is not endowed with a dynamics, namely when $\dot{R}=0$, or $R(\tau)=R$=constant. This case is very important since it tests the consistency of the system of equations obtained previously. In addition, it gives us the requisites for having static glues of slowly rotating spacetimes.
For this case, Eq.~(\ref{sphersymcond}) gives us that $\dot{T}=e^{-\frac{\nu}{2}}$. It is not difficult to show that the solution to the
above system of equations is
\begin{eqnarray}
\epsilon^2B_2^{+}&=&\frac{\epsilon^2_+(h_2^+ + \frac{\partial \nu^{-}}{\partial R_0}k_2^+R_0) - \epsilon^2_-(h_2^- + \frac{\partial \nu^{-}}{\partial R_0}k_2^-R_0) }{\left[\frac{\partial \nu}{\partial R_0}\right]^+_-}\label{B2pR0},\\
\epsilon^2B_2^{-}&=&\frac{\epsilon^2_+(h_2^+ + \frac{\partial \nu^{+}}{\partial R_0}k_2^+R_0) - \epsilon^2_-(h_2^- + \frac{\partial \nu^{+}}{\partial R_0}k_2^-R_0) }{\left[\frac{\partial \nu}{\partial R_0}\right]^+_-}\label{B2mR0},\\
A_2^{\pm}&=&0\label{A2pmR0},\\
\epsilon^2A_{0}^{\pm}&=&-\epsilon^2_{\pm}\left(h_0^{\pm}+\frac{B_0^{\pm}}{2}\frac{\partial \nu^{\pm}}{\partial R}\right)e^{-\frac{\nu^{\pm}}{2}}\tilde\tau\label{A0pmR0},
\end{eqnarray}
with $B_0$ a jump-free ($B_0^+=B_0^-$) and arbitrary constant.
In this case, the induced metric, Eq.~(\ref{inducmetricfinal}), does not depend on $\tilde\tau$, as expected. It can be checked that this is the only case where such an aspect rises concomitantly with the self-consistency of the system of equations. 
This justifies the choice of Eqs.~(\ref{A0cond}--\ref{alpha2}).

It is worth emphasizing that $R(\tilde\tau)$ = constant $\equiv R$ does not automatically guarantee the thin shell's stability upon perturbations. This is just the case whenever $\Sigma$ is bound. From Eq.~(\ref{hypersurfeq}), one sees that stability will be ensured just when both $R(\tilde\tau)$ and $B(\tilde\tau,\tilde\varphi)$ are bound functions. In particular, for our analysis to be meaningful, that should be fulfilled by the latter function. After meeting this demand, the only one left is the bound nature of $R(\tilde\tau)$. This can be analyzed within the thin-shell formalism in the spherically symmetric case, and we know that it is summarized by the search of minima of an effective potential \cite{2005CQGra..22.4869L}. As a result, in the stable case, $\dot{R}$ is automatically bound and could be approximated by a harmonic function around $R$ with a very small amplitude. It is simple to show in this case that the corrections to Eqs. (\ref{B2pR0}--\ref{A0pmR0}) coming from our field equation (\ref{A0cond}--\ref{alpha2cond}) will also be proportional to oscillating functions around $r=R$. More specifically, if $\delta R\equiv R(\tilde{\tau})-R= {\cal A}\cos(\tilde{\omega} \tilde{\tau})$, where ${\cal A}/R\ll 1$ and $\tilde{\omega}$ is a constant, then $\delta A_0\propto \sin(\tilde{\omega} \tilde{\tau})$, $\delta A_2\propto \sin(\tilde{\omega} \tilde{\tau})$, $\delta B_0\propto \cos(\tilde{\omega} \tilde{\tau})$ and $\delta B_2\propto \cos(\tilde{\omega} \tilde{\tau})$. This must be the case mathematically since we have a system of inhomogeneous trigonometric functions to solve. From Eqs. \eqref{hypersurfeq} and \eqref{coordtrans}, it also makes physical sense that $\delta A_0$ and $\delta A_2$ ($\delta B_0$ and $\delta B_2$) oscillate in the same way. Thus, we conclude that the stability of $\Sigma$ in the spherically symmetric case also implies its stability in the presence of small rotations. This is a very important result and will be used in subsequent sections when we elaborate upon the energy conditions of hypersurfaces in the case of slow rotation.

\section{Energy-momentum tensor for a slowly rotating thin shell}
\label{sigmarot}

We now determine the energy-momentum tensor on $\Sigma$ that guarantees the glue of two slowly rotating spacetimes whose boundary is dynamic. 
We must find the normal vector to $\Sigma$ to do so. 
Generally, the normal vector to a given hypersurface $\Psi$ is \cite{2004reto.book.....P}
\begin{equation}
n_{\mu}\equiv \frac{\epsilon_n\partial_{\mu}\Psi}{|g^{\alpha\beta}\partial_{\alpha}\Psi\partial_{\beta}\Psi|^{\frac{1}{2}}}\label{normalSigma},
\end{equation}
where $\epsilon_n=\pm 1$, depending on $\Sigma$ being space-like or time-like, respectively. In addition, Eq.~(\ref{normalSigma}) ensures that $n_{\mu}$ is a unit vector in the direction of growth of $\Sigma$. 

Let us now calculate the gradient to $\Sigma$, $\partial_{\mu}\Psi$. From Eq.~(\ref{hypersurfeq}),
\begin{equation}
\partial_{\mu}\Psi= \left(-\frac{1}{\dot{t}}[\dot{R}+\epsilon^2\dot{B}],1,-\frac{\partial B}{\partial \tilde{\theta}},0\right)\label{partialpsi}.
\end{equation}
From Eqs.~(\ref{slowrotmetric}), (\ref{coordtrans}), (\ref{normalSigma}) and (\ref{partialpsi}), one obtains
\begin{eqnarray}
n_0&=&e^{\frac{\nu+\lambda}{2}}\dot{R}\left\{1 - \epsilon^2\left[\frac{\dot{B}}{\dot{R}} + {e^{\lambda}\dot{R}^2}\left(\frac{\dot{B}}{\dot{R}}-\frac{B}{2}\frac{\partial \nu}{\partial R}-h \right) \right.\right. \nonumber\\ &&\left.\left.- {e^{\nu}\dot{T}^2}\left( \frac{\dot{A}}{\dot{T}}-\frac{B}{2}\frac{\partial \lambda}{\partial R}-j \right) \right] \right\}\label{n0},\\
n_1&=&-e^{\frac{\nu+\lambda}{2}}\dot{T}\left\{1 + \epsilon^2\left[\frac{\dot{A}}{\dot{T}}+ {e^{\lambda}\dot{R}^2}\left(\frac{\dot{B}}{\dot{R}}-\frac{B}{2}\frac{\partial \nu}{\partial R}-h \right)\right.\right. \nonumber\\ &&\left.\left. - {e^{\nu}\dot{T}^2}\left( \frac{\dot{A}}{\dot{T}}-\frac{B}{2}\frac{\partial \lambda}{\partial R}-j \right)\right] \right\}\label{n1},\\
n_2&=&e^{\frac{\nu+\lambda}{2}}\dot{T} \epsilon^2\frac{\partial B}{\partial \tilde{\theta}}\label{n2},\\
n_3 &=& 0.
\end{eqnarray}
In the above normal components, we have assumed that $\Sigma$ is time-like, {thus $\epsilon_n=-1$ in Eq.~(\ref{normalSigma})}. The contravariant components of the normal vector can be worked out simply using $n^{\mu}=g^{\mu\nu}n_{\nu}$.

Now we proceed with the calculations of the extrinsic curvature, defined as \cite{2004reto.book.....P}
\begin{equation}
K_{ab}\equiv n_{\mu;\nu}\frac{\partial x^{\mu}}{\partial y^a} \frac{\partial x^{\nu}}{\partial y^b}= -n_{\mu}\left(\frac{\partial^2x^{\mu}}{\partial y^a\partial y^b}+\Gamma^{\mu}_{\alpha\beta}\frac{\partial x^{\alpha}}{\partial y^a} \frac{\partial x^{\beta}}{\partial y^b}\right)\label{extrinsiccurv}.
\end{equation}
For slowly rotating spacetimes, off-diagonal terms in $K_{ab}$ appear. First, we recall that in the spherically symmetric case
\begin{equation}
K^{0}_{0}=\frac{\nu'(e^{-\lambda}+\dot{R}^2)+2\ddot{R}+\lambda'\dot{R}^2}{2\sqrt{e^{-\lambda}+\dot{R}^2}}\label{K00ss},
\end{equation}
and
\begin{equation}
K^{1}_{1}= K^{2}_{{}2}=\frac{\sqrt{e^{-\lambda}+\dot{R}^2}}{R}\label{K11ss}.
\end{equation}
For the first order corrections in ``$\epsilon$'', we also have
\begin{equation}
K_{2}^{0}=\frac{1}{2}\frac{\partial\omega}{\partial R} R^2\sin^2\tilde{\theta}e^{-\frac{\nu+\lambda}{2}}= -R^2\sin^2\tilde{\theta} K_{0}^{2}\label{K02}.
\end{equation}
According to Lanczos' equation \cite{1924AnP...379..518L,1966NCimB..44....1I}, the surface energy-momentum leading to a distributional solution to Einstein's equations is
\begin{equation}
8\pi S^a_b= [K^a_b]^+_- - \delta^a_b [K]^+_-\label{Sabdef},
\end{equation}
where $K\doteq h^{ab}K_{ab}$ is the trace of the extrinsic curvature.

In the spherically symmetric ($ss$) case, $S^a_b$ is a diagonal tensor concerning the $(\tilde\tau,\tilde\theta,\tilde\varphi)$ coordinates, i.e.,
\begin{eqnarray}
S^a_b&=& \frac{1}{8\pi} {\rm diag}(-2[K^1_1]^+_-,-[K^0_0+K^1_1]^+_-,-[K^0_0+ K^1_1]^+_-) \nonumber\\
&\equiv& {\rm diag}(\sigma_{ss},-{\cal P}_{ss}, -{\cal P}_{ss})\label{Sabss}.
\end{eqnarray}
This is the energy-momentum tensor for a comoving frame. Thus, the fluid on $\Sigma$ is perfect-like, i.e.,
\begin{equation}
S^{ab}=\sigma_{ss} u^au^b + {\cal P}_{ss}(u^au^b-h^{ab}) \label{Sabssperfect}.
\end{equation}
Whenever one works up to the first-order approximation in ``$\epsilon$'', similar results as the above ones also hold in the coordinate system $(\tilde\tau,\tilde\theta,\tilde{\varphi})$, with $\tilde{\varphi}$ defined by Eq.~(\ref{coordtrans}).

Now we turn to the case where up to the second order corrections in ``$\epsilon$'' are worked out. From what we have pointed out previously, the form of the surface energy-momentum in this case in the coordinate system $(\tilde\tau, \tilde\theta, \tilde\varphi)$ should generically resemble as
\begin{equation}
S^a_b = \left[
\begin{array}{ccc}
S^0_0 +\epsilon^2\bar{S}^0_0 & \epsilon^2\bar{S}^0_1 & \epsilon\bar{S}^0_2 \\
\epsilon^2\bar{S}^1_0 & S^1_1+\epsilon^2\bar{S}^1_1 & 0 \\
\epsilon\bar{S}^2_0 & 0 & S^1_1+\epsilon^2\bar{S}^2_2 \label{Sabgenerically}
\end{array}
\right].
\end{equation}
Because we are just inserting rotation into the problem, we still expect the fluid on $\Sigma$ to be perfect-like. Thus, from Eq.~(\ref{Sabssperfect}), we have that
\begin{equation}
S^a_b u^b=\sigma u^a\label{eigenvalueeq}.
\end{equation}
%
The above equation clearly tells us that $\sigma$ is an invariant and hence it will be the same for every coordinate system on $\Sigma$. The results coming from lower orders of approximation tell us that we should accept solutions of the form
\begin{equation}
u^a= [u_0+\epsilon^2\bar{u}_0,\epsilon^2\bar{u}_1,\epsilon\bar{u}_2],\;\; \mbox{and}\;\; \sigma=\sigma_{ss}+\epsilon^2\bar{\sigma} \label{velsigmaa2}.
\end{equation}
It can be shown that the only solution coming from Eq.~(\ref{eigenvalueeq}) that satisfies such prerequisites is
\begin{eqnarray}
u^a&=&\Bigg \{1 +\epsilon^2\left[ \frac{1}{2}\left(\frac{R\sin\tilde\theta \bar{S}^2_0}{S^0_0-S_1^1}\right)^2 - \{ \alpha_2 \}^+_-  P_2(\cos\tilde\theta)\right], \nonumber\\ && \frac{\epsilon^2\bar{S}^1_0}{S^0_0-S_1^1},
\frac{\epsilon\bar{S}^2_0}{S^0_0-S_1^1}\Bigg \}\label{vela2},
\end{eqnarray}
where we fixed the arbitrary quantities $u_0$ and $\bar{u}_0$ coming from the eigenvalue approach by the normalization condition $u^au_a=1$, and
\begin{equation}
\sigma=S^0_0+\epsilon^2\left(\bar{S}^0_0 + \frac{\bar{S}^0_2\bar{S}^2_0}{S^0_0-S_1^1} \right)\label{sigmaa2}.
\end{equation}
For the (invariant) surface tension, from Eq.~\eqref{eigenvalueeq}, we generically have that
\begin{equation}
{\cal P}=\frac{1}{2}S^a_b(u^bu_a-\delta^b_a)\label{pressuregen}.
\end{equation}
For the slowly rotating thin-shell case, the above equation becomes
\begin{equation}
{\cal P}= -S^1_1 +\epsilon^2\left[-\frac{\bar{S}^1_1+\bar{S}^2_2}{2} + \frac{\bar{S}^0_2\bar{S}^2_0}{2(S^0_0-S_1^1)} \right]\label{pressurea2}.
\end{equation}

Note that the second order correction in ``$\epsilon$'' to ${\cal P}$ is polar angle-dependent, and it is different from the one to $\sigma$ [see Eqs.~(\ref{K02}), (\ref{Sabdef}), (\ref{sigmaa2}) and (\ref{pressurea2})]. We finally stress that our perturbative analyses break down whenever the surface quantities in the spherically symmetric are null. That is natural because we assume the slow-rotation case is only a small correction to the spherically symmetric case.  

%
%

\section{Dragging of inertial frames}
\label{drag}

Here we briefly discuss some kinematic effects related to the match of two slowly rotating Hartle's spacetimes [see Eq.~(\ref{slowrotmetric})].  Let us first analyze aspects of a static observer inside the rotating thin shell. This observer can be described by $d\varphi^-=0$, or from Eq.~(\ref{coordtrans}), equivalently $d\tilde\varphi= -(\epsilon C)^-d\tilde\tau$. Thus, in the fixed stars' frame of reference (stationary observers at infinity), such an observer is rotating with the angular velocity
\begin{equation}
\frac{d\varphi^+_{(d\varphi^-=0)}}{dt^{+}}= [(\epsilon C)^+-(\epsilon C)^-]\frac{d\tilde\tau}{dt^+}= (\epsilon\omega)^+-\frac{(\epsilon\omega \dot T)^-}{\dot{T}^+}\label{angvelintobsfs},
\end{equation}
where we have used also Eq.~(\ref{cvalue}). We recall that the $\dot{T}^{\pm}$ can easily be read off from Eq.~(\ref{sphersymcond}). Equation (\ref{angvelintobsfs}) states that far away external observers see internal ones rotating, even when $\epsilon^-=0$, namely when the inner spacetime is spherically symmetric. Such an effect is the well-known dragging of inertial frames, or Lense-Thirring effect \cite{1973grav.book.....M}. The case where $\epsilon^{-}=0$ is interesting because it shows the rotation of the shell intrinsically induces a rotation of observers inside it. Whenever the inner spacetime is also endowed with a rotational parameter ($\epsilon^-\neq 0$), naturally, it also contributes to the final angular velocity fixed stars ascribe to internal observers at rest, as evidenced by Eq.~(\ref{angvelintobsfs}). Even more remarkable in this case is the possibility of the disappearance of the dragging of inertial frames [see Eq.~(\ref{angvelintobsfs})]. This is so even when $\epsilon^{\pm}$ and the shell parameters are given by convenient values of the inner spacetime parameters.

Also, about distant observers, the rotation of the thin shell can be obtained. We know that concerning the frame with coordinates
$(\tilde\tau, \tilde\theta,\tilde\varphi)$, the shell rotates with {angular} velocity
\begin{equation}
\frac{d\tilde\varphi}{d\tilde\tau}= \frac{\epsilon\bar{S}^2_0}{S^0_0-S^1_1}\label{shellrottildesyst},
\end{equation}
where we recall that the ``$\epsilon$'' dependence on Eq.~(\ref{vela2}) is merely an indicator that the associated surface energy-momentum tensor is
of a given order on the rotational parameters $\epsilon^{\pm}$. 
Taking into account Eqs.~(\ref{shellrottildesyst}) and (\ref{coordtrans}), we have that the angular velocity of the shell relative to the fixed stars is
\begin{equation}
\frac{d\varphi^+_{\Sigma}}{dt^+}= \frac{d\tilde\varphi}{dt^+}+\epsilon^+C^+\frac{d\tilde\tau}{dt^+}= \frac{\epsilon\bar{S}^2_0}{(S^0_0-S^1_1)\dot{T}^+}+(\epsilon\omega)^+\label{velSigmafs},
\end{equation}
where also Eq.~(\ref{cvalue}) has been taken into account and we recall that $\dot{T}$ is given by Eq.~(\ref{sphersymcond}). For completeness,
we remark that the relative velocity of observers inside the shell with the shell itself could be obtained out of Eqs.~(\ref{angvelintobsfs}) and (\ref{velSigmafs}) and always vanishes at the associated ``gravitational radius'' of the external spacetime ($e^{\nu_+}=0$), showing that rigid dragging takes place in this case \cite{1968PhRv..170.1187D}. Nevertheless, whenever a nontrivial inner spacetime is considered, in principle, configurations always exist that would lead the observers inside the shell to corotate with the thin shell. The interest in the effect of frame dragging \cite{2011PhRvL.106v1101E,2020Sci...367..577V} naturally lies in the fact that its measurement could give direct information about the stratification of spacetime.

\section{Matching slowly rotating Kerr spacetimes}
\label{KN}

A case of physical interest concerning the glue of slowly rotating spacetimes would be where they are of Kerr types. This could be, for instance, a model for slowly rotating neutral thin shells of matter collapsing onto Kerr black holes as it could happen in AGNs.
A natural advantage of this match is the geometric simplicity of both regions. We know that the Kerr metric in the Boyer-Lindquist
coordinates ($\bar t, \bar r,\bar \theta,\bar\varphi$) is \cite{1973grav.book.....M}
\begin{eqnarray}
ds^2&=& \left(1- \frac{2M\bar{r}}{\rho^2}\right)d\bar{t}^2 - \frac{4M\bar{r}a\sin^2\bar{\theta}}{\rho^2}d\bar{t}d\bar{\varphi}\nonumber\\ &-&\frac{\rho^2}{\Delta}d\bar{r}^2- \rho^2d\bar{\theta}^2-\frac{\Upsilon}{\rho^2}\sin^2\bar{\theta} d\bar{\varphi}^2\label{kerrnewman},
\end{eqnarray}
where
\begin{eqnarray}
\Delta&\equiv& \bar{r}^2-2M\bar{r}+a^2\label{Deltakn},\\
\Upsilon &\equiv& (\bar{r}^2+a^2)^2 - \Delta a^2\sin^2\bar{\theta},\label{varphi2factorkn}\\
\rho^2&\equiv& \bar{r}^2+a^2\cos^2\bar{\theta}\label{rhokn}.
\end{eqnarray}
This solution has two arbitrary constants: the system's total mass, $M$, and its total angular momentum per unit mass, $a$. The above metric has horizons at $\Delta=0$. We shall not elaborate any more on the well-known properties of this solution.

We emphasize that Hartle's coordinates ($t,r,\theta,\varphi$) differ from the Boyer-Lindquist ones. It is simple, though tedious, to show that the coordinate
transformations linking the aforesaid coordinate systems up to second order in ``$\epsilon$'' (here $\epsilon = a/r$) are (see, e.g., \citet{2022PhRvD.105b4002B})
%
\begin{eqnarray}
\bar{\theta}&=&\theta -\frac{a^2}{2r^2}\left(1+\frac{2M}{r} \right)\cos\theta\sin\theta, \label{thkntothht}\\
\bar{r}&=& r  -\frac{a^2}{2r}\left[ \left(1+\frac{2M}{r} \right) \left(1-\frac{M}{r} \right) \right. \nonumber\\ &-&\left. \left(1-\frac{2M}{r} \right) \left(1+\frac{3M}{r}\right)\cos^2\theta \right] \label{rkntorht}.
\end{eqnarray}
%

The coordinate transformations given by Eqs.~(\ref{thkntothht}) and (\ref{rkntorht}) are necessary to apply the formalism we developed previously for matching two slowly rotating spacetimes. The Kerr metric components up to second order in ``$\epsilon$'' in Hartle's coordinates are
\begin{align}
&h_0^{K}(r)=\frac{M^2}{r^2}e^{-\nu},\label{h0kn}\\
&h_{2}^{K}(r) = \frac{M}{r^3}(r-2M)(M+r)e^{-\nu}\label{h2kn},\\
&j_0^{K}(r) =  -\frac{M^2}{(r-2M)^2} e^{\nu}\label{j0kn},\\
&j_2^{K}(r) = \frac{M (5 M - r)}{r(r-2M)}e^{\nu}\label{j2kn},\\
&k_2^{K}(r) = -\frac{M (2 M+r)}{r^2}\label{k2kn},
\end{align}
and
\begin{align}
e^{\nu} &= 1-\frac{2M}{r},\label{ealpha}\\
\omega &= \frac{2 M }{r^2}\label{omegakn}.
\end{align}

We now investigate the induced shell quantities (surface energy density and tension) due to small rotations. 
For example, let us choose thin shells in equilibrium at the radial coordinate $R=5M_+$, $M_+$ being the mass of the outer Schwarzschild spacetime seed. This analysis is interesting since, in principle, test particles cannot be in stable equilibrium for $r<6M_+$ in a Schwarzschild spacetime, unline thin shells (which renders them special and shows the importance of searching for distributional solutions in general relativity). In addition, it would allow us to probe thin shells in strong gravitational fields, as motivated by inner accretion discs in AGNs \cite{2012MNRAS.424..217F} or even in stratified neutron stars. In what follows, for numerical convenience, we take $a_-= n a_+$ and $M_-= m M_+$, with $(n,m) \in (\Re, \Re^+)$, respectively. As expected, for two Schwarzschild spacetimes, when $M_+>M_-$, all induced surface quantities are positive \cite{2005CQGra..22.4869L}.

To proceed, we must also choose other parameters related to the glued thin shells. Our analysis in section \ref{Rconst} concluded that shells in equilibrium could be characterized by an arbitrary jump-free constant $\epsilon^2B_0$, related to the effective change of radius from $R$ due to rotation. 
We choose for our investigations the case where $B_0=0$. The reasons for that are mainly connected with its physical reasonableness. One can verify that the glue of two Kerr-spacetimes at $R=5M_+$ is such that for reasonable $n$ (such as $-1<n<1$) and $m<1$, one has that $B(\theta =0)<0$ and $B(\theta=\pi/2)>0$, meaning that the resultant surface is flattened in the poles and stretched in the equator with respect to a sphere, as one expects physically when rotation takes place. 
Situations where $B_0\neq 0$ are left to be investigated elsewhere.

As we have shown previously, the stability analyses of slowly rotating thin shells (around given radial positions) can be
summarized by their spherically symmetric counterparts [see section \ref{Rconst}]. Therefore, we should analyze the stability of glued Schwarzschild spacetimes to investigate slowly rotating Kerr spacetimes. It is known in this case that a thin shell is stable at $r=R$
\textit{iff} its associated effective potential \cite{2005CQGra..22.4869L}
\begin{equation}
V(R)\equiv \frac{1}{2}(e^{-\lambda_-}+e^{-\lambda_+})-\frac{1}{4}(4\pi R \sigma_{ss})^2-\frac{1}{4}\left(\frac{[e^{-\lambda}]^+_-}{4\pi R \sigma_{ss}} \right)^2\label{effpot}
\end{equation}
has a minimum there ($d^2V/dR^2>0$). It can be shown that \cite{PhysRevD.90.123011}
\begin{equation}
\frac{d^2V}{dR^2}= -\frac{\left [e^{-\frac{\lambda}{2}} \left\{(2\eta +1)\left(1+\frac{R}{2}\lambda'\right)-\frac{R^2}{2}\left(\nu''-\frac{1}{2}\lambda'\nu'\right)\right\}\right]^+_-}{R^2[e^{\frac{\lambda}{2}}]^+_-}, \label{ddV}
\end{equation}
where $\eta$ is defined as the adiabatic speed of sound squared on the shell \cite{PhysRevD.90.123011}, i.e., $\eta\equiv c_{s}^2$. We stress that for the thin shell formalism to make sense, $\sigma_{ss}\neq 0$, which we take in our analysis. That also justifies the perturbative treatment with respect to the spherically symmetric case in the previous sections.

\begin{figure}[!htbp]
\centering
\includegraphics[width=\hsize,clip]{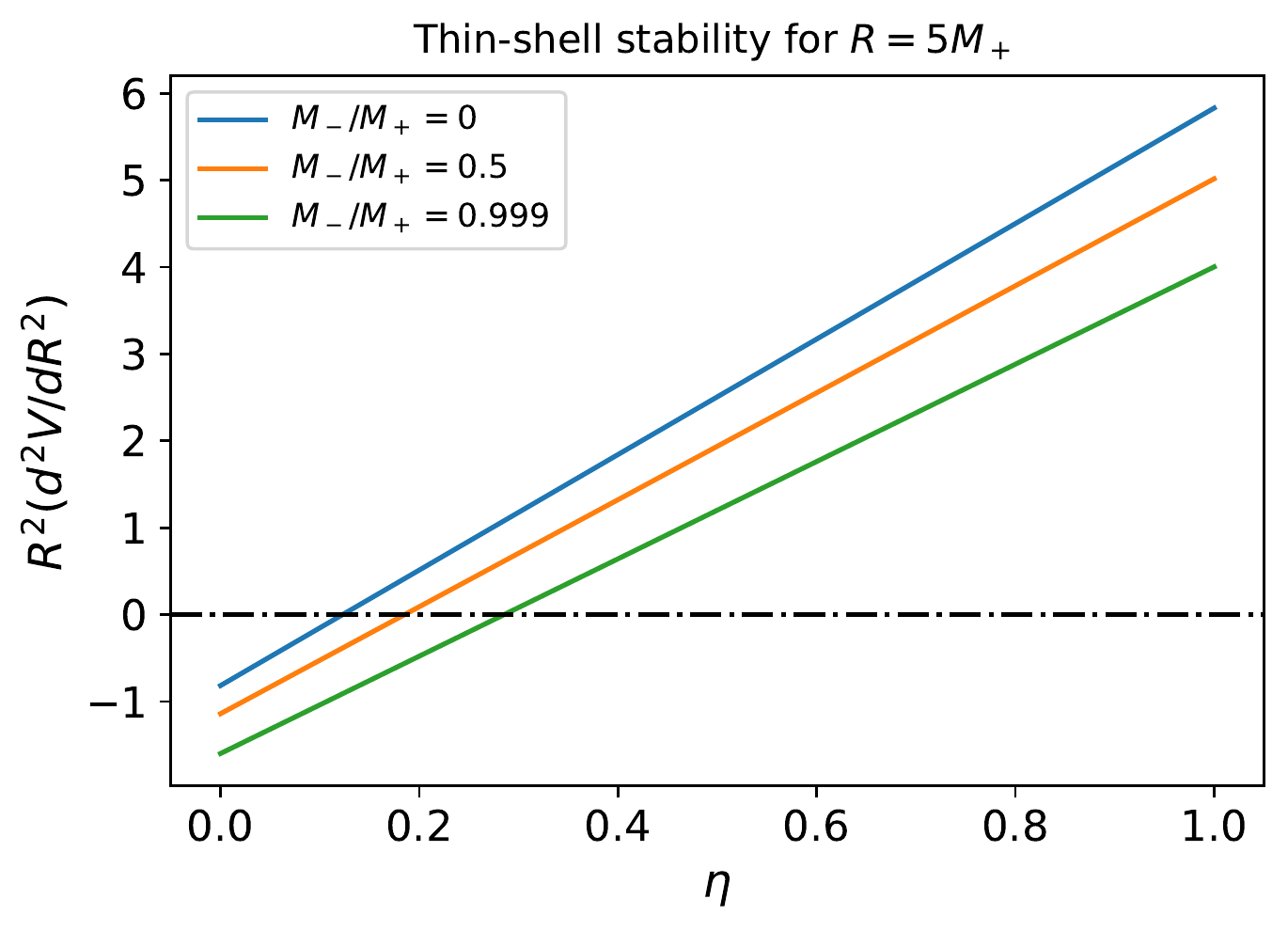}
\caption{Thin-shell stability for some glues of Schwarzschild spacetimes where the outer masses are always larger than the inner ones, leading to $\sigma_{ss}>0$ and ${\cal P}_{ss}>0$. The equilibrium radius is taken $R=5M_+$ for all cases. For Schwarszchild spacetimes, the quantity of relevance is $R/M_+$, and $M_+$ could be any. In particular, when $R/M_+=5$, several thin-shells' speeds of sound lead to stable equilibria. That is the case even for very small thin-shell masses ($M_-/M_+\sim 1$). For the case $M_-/M_+=0.999$, any $\eta \gtrsim 0.3$ will fulfill the stability condition.}
\label{stabss}
\end{figure}

Figure \ref{stabss} shows the stability condition ($R^2d^2V/dR^2>0$) for several matches of Schwarzschild spacetimes with masses leading to $\sigma_{ss}>0$ at $R=5M_+$. One sees from the aforementioned plot that thin shells are stable for most $\eta$ and internal-to-external mass ratios. This suggests that $r=5M_+$ is a feasible choice for the equilibrium radius of a thin shell with nontrivial surface degrees of freedom. 

\begin{figure}[!htbp]
\centering
\includegraphics[width=\hsize,clip]{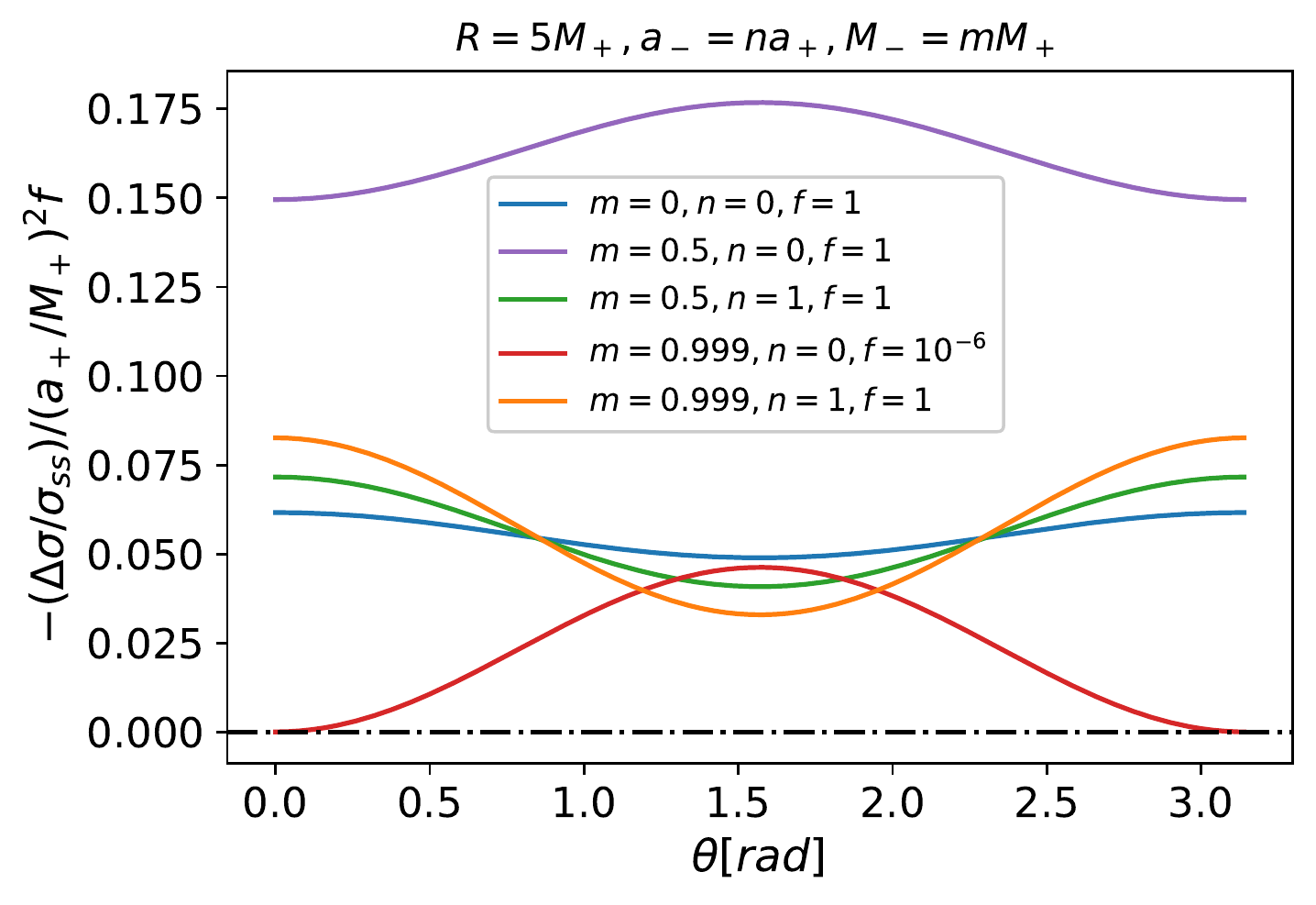}
\caption{Relative surface energy density corrections due to rotation ($\Delta\sigma/\sigma_{ss} \equiv  \sigma /\sigma_{ss}-1$) for some parameters of the glued spacetimes. Internal and external mass ratios were chosen to coincide with those of Fig.  (\ref{Deltasigma}). To be more general, we did not specify the particular rotational parameter $(a_+/M_+)^2$; we just assumed it was small. The factor ``$f$'' is only for convenience (plotting all the cases in a similar scale). Negative $n$'s lead to the same results as their positive counterparts. Note that for all the cases and polar angles, $\Delta\sigma <0$, meaning that rotation decreases the local surface energy density on the thin shell.}
\label{Deltasigma}
\end{figure}

\begin{figure}[!htbp]
\centering
\includegraphics[width=\hsize,clip]{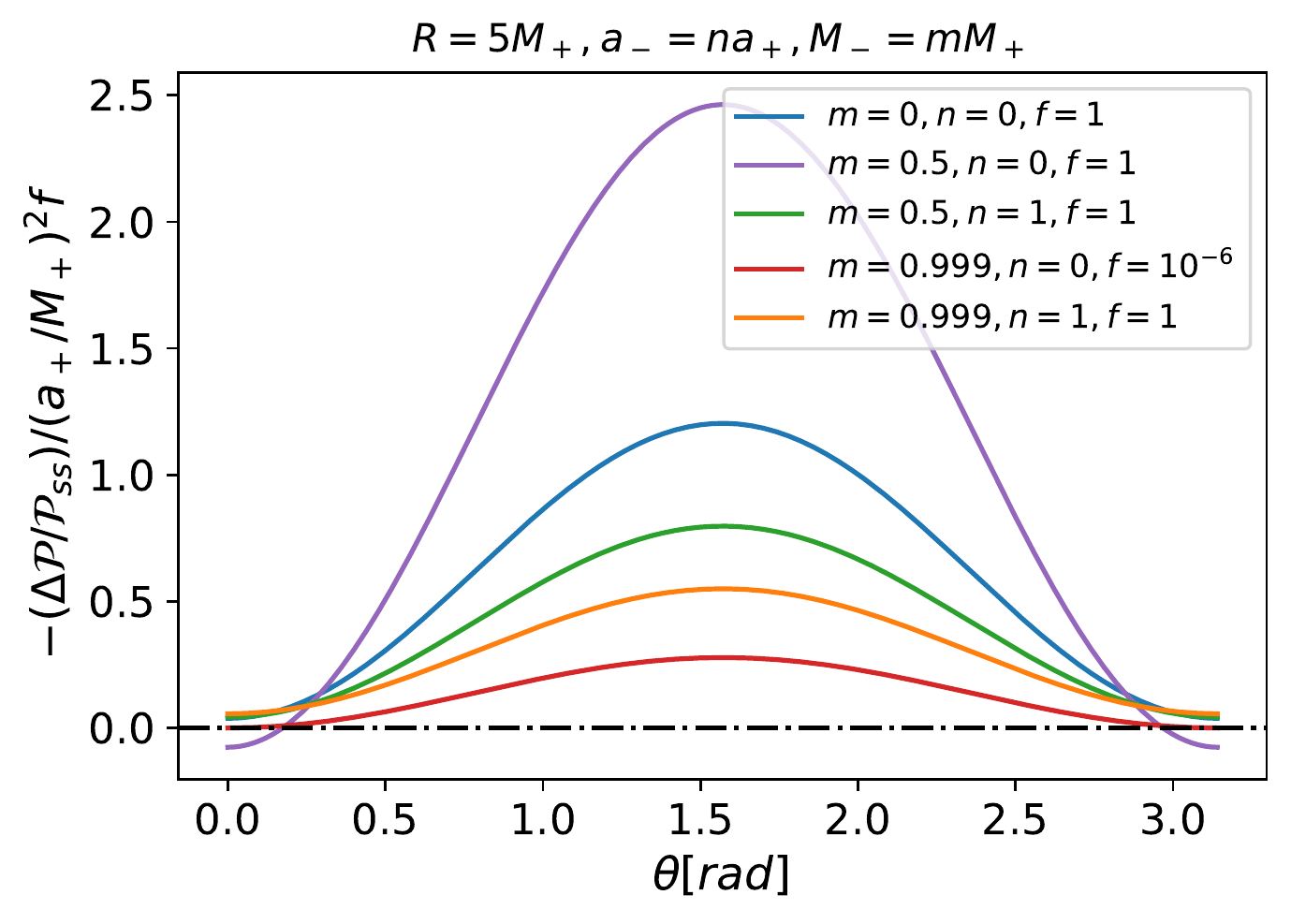}
\caption{Relative induced surface tension ($\Delta{\cal P}/{\cal P}_{ss} \equiv {\cal P}/{\cal P}_{ss}-1$) as a function of the polar angle $\theta$. We have chosen the same mass ratios of Fig.  (\ref{Deltasigma}). Note that for almost all choices of thin-shell mass and polar angles, the surface tension correction to the spherically symmetric case due to rotation is negative.}
\label{Deltap}
\end{figure}

We study induced surface quantities on slowly rotating neutral thin shells in slowly rotating Kerr spacetimes. Figure \ref{Deltasigma} shows the relative changes for the surface energy densities ($\Delta \sigma/\sigma$) for some selected Schwarzschild seeds. For all polar angles, they decrease when compared to their spherically symmetric counterparts. That is the case if the inner spacetime counter-rotates with respect to the outer one (because $\Delta \sigma$ with positive and negative $n$ do not differ much), if it has no rotation at all ($n=0$), or if it has the same rotation as the external one ($n=1$). That is the same concerning relative changes of the relative surface tension $\Delta{\cal P}/{\cal P} \equiv {\cal P}/{\cal P}_{ss}-1$ (see Fig. \ref{Deltap}). 

\begin{figure}[!htbp]
\centering
\includegraphics[width=\hsize,clip]{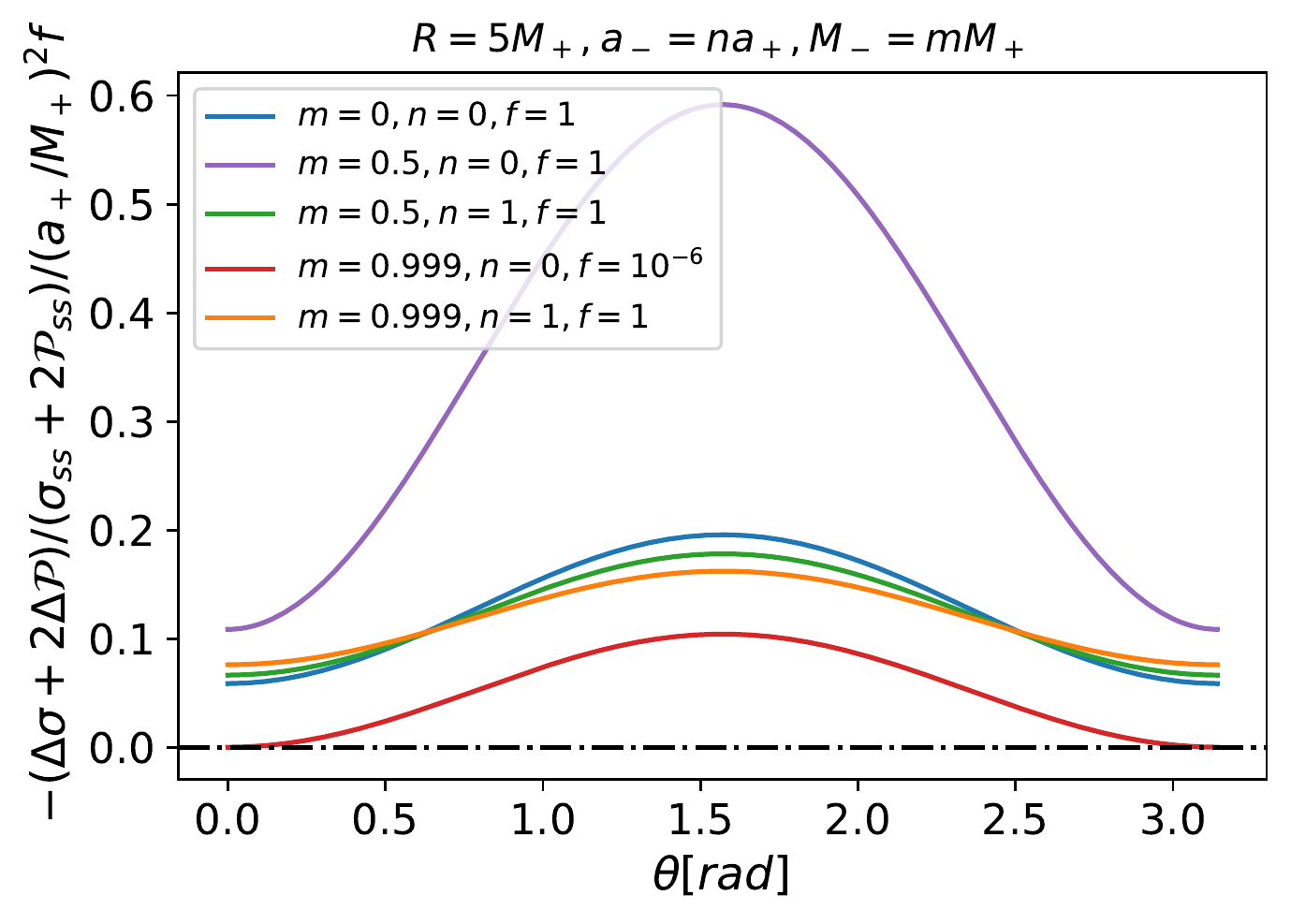}
\caption{ The parameters for the curves are the same as their counterparts in Fig.~(\ref{Deltasigma}). The induced part due to the rotation in the strong energy condition ($\Delta \sigma + 2 \Delta {\cal P}$) is negative (clearly not the case for the selected values of the glued spacetimes, whose spherically symmetric counterparts satisfy all energy conditions), which points to the possibility of the non-validity of some energy conditions in non-perturbative analyzes.}
\label{sec}
\end{figure}

For all the cases in Fig. \ref{stabss}, it is simple to verify that the weak, null, strong, and dominant energy conditions \cite{2004reto.book.....P}
are satisfied. This shows the reasonableness of our assumptions. In the presence of rotation, the aforesaid conditions are slightly weakened. Fig. \ref{sec} exemplifies the previous statement due to the negativity of the induced part of the strong energy condition ($\Delta\sigma + 2\Delta{\cal P}< 0$) for the spacetime matches in Fig. \ref{stabss}. Similar conclusions could be reached for the weakening of other energy conditions (weak and null) because both $\Delta \sigma$ and $\Delta {\cal P}$ are negative. Naturally, in the perturbative scope, this does not mean any violation of the energy conditions but only their weakening when small rotations are present, which suggests nontrivial behavior in the non-perturbative case.

\section{Concluding remarks}
\label{conc}

This work showed the subtleties and nontrivialities in matching two slowly rotating Hartle's spacetimes through dynamic hypersurfaces. For a given matching hypersurface, we have obtained generically that its equilibrium points are stable if their spherically symmetric counterparts are so. Concerning the kinematical effects, we have shown that it is possible to match spacetimes where a rigid-rotation behavior appears--at least in some limit--and the frame-dragging effect can give information about the matched spacetimes. We also found that the thin-shell's surface energy densities and surface tensions decrease compared to their spherical counterparts. This suggests, for instance, that the assumption of having everywhere positive surface degrees of freedom may be broken in non-perturbative calculations. Energy conditions may also be violated in this scenario, which could have important consequences (for the relevance of the energy conditions in general relativity, see Chapter 34 of \cite{1973grav.book.....M} and \cite{2020CQGra..37s3001K}). Another possibility would be that the possible violation of the energy conditions would preclude the existence of such thin-shell structures. Given its importance, we plan to investigate that better elsewhere.

Let us elaborate on some of the above points. The automatic stability of a slowly rotating thin shell, when its spherically symmetric seed is stable, is reasonable because a slowly rotating spacetime and thin shell perturbations there can be roughly seen as effective thin shell perturbations in the spherical case. The fact that this stability emerges from our system of equations when matching two slowly rotating spacetimes is also relevant because it strengthens its consistency and it shows that when the thin shell stability is concerned, it suffices to analyze just the spherically symmetric case. On the other hand, the decrease in the value of surface quantities when the slow rotation is present (as was clear when matching two Kerr spacetimes) is not trivial. A possible interpretation of it is that part of the thin-shell energy goes into rotational energy. That would reinforce the need, in general, to go beyond the slow rotation approximation we have adopted. Clearly, this is not an easy task because some symmetries of the spacetimes to be matched are lost. However, it is worth mentioning that, for instance, in the case of rotating neutron stars, deviations from the spherical symmetry are essentially negligible for rotation rates of up to a few hundred Hz (see, e.g., \cite{2014NuPhA.921...33B, 2015PhRvD..92b3007C}). Thus, we expect the slow rotation Hartle spacetime metric to be a reasonable approximation for those configurations. Finally, Figs. \ref{Deltasigma}, \ref{Deltap} and \ref{sec} also show that the largest relative differences for the surface tension and surface energy density when $m$ is not too close to the unity happen when the internal spacetime is flat. The reason here is that this case leads to the largest mass-energy content of the thin shell. Thus, one would also expect that their relative differences would be maximized with respect to other cases where the internal spacetime has mass and rotation. The results for $m$ close to one should be taken with a grain of salt because the perturbative model investigated breaks down when $m=1$ (no surface degrees of freedom are induced in the background spacetime).

One could apply the thin shell formalism developed here for cases involving astrophysical black holes, which are believed to be of Kerr type. That could be for those at the centers of galaxies, such as AGNs, or even those in binaries or alone. The idea is that structures (thin shells) could be formed around black holes, wrapping them up totally or partially. For the case of black holes with masses ranging from one to two solar masses and thin shells with a (small) fraction of that mass, if their equilibrium position is $R=4$--$6 M_{+}$, then $R$ is of the order of a neutron star's radius ($\approx 12$ km for a $1.4$ $M_{\odot}$ star \cite{2019ApJ...887L..21R,2019ApJ...887L..24M}). The intriguing question is whether external observers could perceive those thin shells as neutron stars. If, instead, BHs of around a solar mass are wrapped up by thin shells and equilibrium radii around $R=(10^3$--$10^4)M_+$, one could perceive them as white dwarfs.
In the era of gravitational-wave astronomy, these possibilities seem interesting to be better investigated because the above equilibrium radii are stable for a large range of sound speeds on the thin shells. Some gravitational-wave aspects are already known for some exotic compact objects (see, e.g., \citep{2015PhRvD..92l4030P,2016PhRvD..94f4015U,2017PhRvD..96b3005M,2017PhRvD..95h4014C,2019PhRvD..99j4050R,2018PhRvL.120h1101M,2019LRR....22....4C,2020PhRvD.102l3010J,2021PhRvD.104h4056N,2021hgwa.bookE..29M}).

Another special arena of application of this work is stratified compact stars, which have in their interiors a huge range of densities and pressures and even different matter phases (e.g., solid and liquid hadronic phases \cite{HaenselPY2007} or possibly even quark and hadronic phases--hybrid stars \cite{2018ApJ...860...12P}). Slow rotation would be the natural extension of the spherically symmetric case, where the stability formalism for stratified stars is already known \cite{2015ApJ...801...19P} and surface degrees of freedom can be induced upon perturbations \cite{2019ApJ...871...47P}. 
The approach presented here can be applied to the match of various matter phases since the second-order slow rotation approximation is accurate for the description of uniformly rotating neutron stars up to frequencies $\sim 300$~Hz (see, e.g., Refs.~\cite{2005PhRvD..72d4028B,2014NuPhA.921...33B}). In this line, comparing and contrasting the stability when a thin shell matches different phases for static and rotating configurations seem relevant. The formalism could also be applied if the surface of a neutron star is buried with a thin layer of supernova debris (e.g., as the so-called central compact objects--CCOs \cite{2011MNRAS.414.2567H,2012MNRAS.425.2487V,2017JPhCS.932a2006D}). In addition, one could check whether or not the stringent stability of strange quark stars with an outer crust obtained in \cite{PhysRevD.90.123011} change when adding rotation. The case of neutron stars rotating in the kHz region needs the non-perturbative solution of the match of axially symmetric spacetimes, which represents the ultimate goal of the analysis proposed in this work.
\begin{acknowledgments}
We are grateful to Jaziel G. Coelho for the useful comments which helped us improve the work. JPP gratefully acknowledges the financial support of FAPES--Fundação do Estado do Espírito Santo--, grant number 04/2022.
\end{acknowledgments}

\bibliographystyle{apsrev4-1}
\bibliography{gluebib}

\end{document}